\documentclass[aps,prd,preprint,groupedaddress,amsfonts,amssymb,amsmath,showpacs]{revtex4}

\usepackage{graphicx}

\begin{document}

\title{Self--similar imploding relativistic shock waves}

\author{J.~C. Hidalgo$^{1,2}$}
\email[Email address: ]{c.hidalgo@qmul.ac.uk}
\author{S. Mendoza$^{1}$}
\email[Email address: ]{sergio@astroscu.unam.mx}
\affiliation{$^1$Instituto de Astronom\'{\i}a, Universidad Nacional
                 Aut\'onoma de M\'exico, AP 70-264, Distrito Federal 04510,
	         M\'exico \\
             $^2$Current address: Astronomy Unit, School of Mathematical 
	         Sciences, Queen Mary, University of London, Mile End Road, 
		 London  E1 4NS, United Kingdom
            }

\date{\today}

\begin{abstract}
  Self--similar solutions to the problem of a strong
imploding relativistic shock wave are calculated.  These solutions
represent the relativistic generalisation of the Newtonian
Gouderley--Landau--Stanyukovich problem of a strong imploding spherical
shock wave converging to a centre.  The solutions are found assuming
that the pre--shocked flow has a uniform density and are accurate for
sufficiently large times after the formation of the shock wave. 
\end{abstract}

\pacs{47.75.+f 47.40.Nm}

\maketitle

\section{Introduction}

  When a large quantity of energy is injected on a very small volume
over an instantaneous  time, matter and energy diverge from the
deposited volume. This phenomenon generates a very high pressure
over the surrounding gas producing an explosion.  When the pressure
difference between the region of explosion and the matter at rest is
sufficiently large, a strong explosion is produced and the discontinuity
gives rise to a divergent shock wave.  An explosive shock wave has a
perfect spherical shape if the volume where the explosion took place
is sufficiently small and there are no anisotropies or barriers in the
surrounding medium.  It was first \citet{sedov46} who formulated and
solved analytically this problem \citep[see also][]{sedov}.  The problem
was also treated by Stanyukovich in his PhD dissertation \citep[see][for a
full description of the general solution]{stanyukovich} and in less detail
by \citet{taylor45}.  Both authors derived the corresponding self--similar
equations and obtained numerical solutions.  As shown by \citet{sedov},
the self--similarity index of the problem is obtained by requiring the
energy inside the shock wave to be a fundamental constant parameter.
Under these circumstances, the famous Sedov--Taylor similarity index
\( \alpha  = 2 / 5 \) is obtained \citep{daufm,zeldovichraizer} for
a similarity variable \( \xi = r / t^{ \alpha} \), where \( r \) is
the radial distance measured from the point of explosion and \( t \)
represents time.

  The inverse physical situation is represented by a strong
imploding spherical shock wave propagating into a uniform
density medium.  This was first investigated by \citet{guderley42}
and fully solved by Landau \& Stanyukovich \citep[see for
example][]{daufm,stanyukovich,zeldovichraizer}.  In this case,
a detailed construction of the self--similar solution was found by
demanding the solution to pass through a singular point admitted by the
hydrodynamical equations.  The similarity variable in this case is given by \(
\eta = r / (-t)^\alpha \) with a similarity index \( \alpha = 0.68837 \) 
for a monoatomic gas.

  The first attempt to build similarity solutions for
relativistic hydrodynamics was introduced by the pioneering work of
\citet{eltgroth71,eltgroth72} with the identification of a self--similar
variable \( J := v / c \), where \( v \) is the velocity of the flow
and \( c \) the speed of light.  Soon after that, \citet{blandford76}
solved the relativistic similarity problem for a very strong explosive
shock wave.  On their analysis, they introduced a similarity variable \(
\chi = \left( 1 -  r /  c t \right) \left[ 1+ 2 ( m + 1 ) \Gamma^2 \right],
\) where \( \Gamma \) represents the Lorentz factor of the explosive
shock front measured in the rest frame of the un--shocked gas and \( m \)
is the similarity index.  Their analysis represents a relativistic
generalisation of the work carried out by Sedov--Taylor--Stanyukovich.

  In this article we construct self--similar solutions for
the case of a very strong  imploding shock wave converging to
a centre.  This constitutes a relativistic generalisation to the
Guderley--Landau--Stanyukovich problem.  In order to obtain a unique
description of the motion, as it happens for the non--relativistic case,
we force the solution to pass through an admissible singularity point,
which in turn fixes the value of the similarity index of the flow.

  The present article is organised as follows.  The next section
describes briefly the main equations for relativistic hydrodynamics on
a flat space--time and mentions the main results for shock waves on any
inertial system of reference.  Later in the article, we write down the
self--similar equations behind the flow of a strong relativistic imploding
shock wave.  We calculate the similarity variable, the self--similar
index of the problem and solve numerically the equations of motion.
Finally, we describe the main properties of the solution.

\section{Shock waves in relativistic hydrodynamics}
\label{hydrodynamics}

 The equations that govern the motion of a  relativistic flow on
a flat space--time are well described by \citet{daufm}.  First, the
conservation of energy and momentum are represented by the condition
that the divergence of the energy--momentum tensor \( T^{\mu\nu} \)
must vanish, i.e.

\begin{equation}
  \frac{ \partial T^{\mu\nu} }{ \partial x^\mu } = 0,
\label{eq.1}
\end{equation}

\noindent with 

\begin{equation}
  T^{\mu\nu} = w u^{\mu} u^\nu - p \eta^{\mu\nu}.
\label{eq.2}
\end{equation}

\noindent In here and in what follows Greek indices have values \( 0,\ 1,\
2,\ 3 \) and Latin ones have values \( 1,\ 2,\ 3 \).  As usual, we use
Einstein's summation convention over repeated indices.  The coordinates
\( x^\alpha \) are such that \( x^\alpha = \left( ct,\ \boldsymbol{r}
\right) \), where \( \boldsymbol{r} \) is the spatial radius vector, \(
t \) represents the time and \( c \) is the velocity of light.  For a
flat space--time, the metric tensor \( \eta_{\mu\nu} \) is given by \(
\eta_{00} = 1 \), \( \eta_{11} = \eta_{22} = \eta_{33} = -1 \) and \(
\eta_{\alpha\beta} = 0 \) for \( \alpha \neq \beta \).  The four--velocity
\( u^\alpha \) is defined by the relation \( u^\alpha = \mathrm{d}
x^\alpha / \mathrm{d} s \), where the interval \( \mathrm{d} s \) is given
by \( \mathrm{d} s^2 = \eta_{\mu\nu} \mathrm{d}x^\mu \mathrm{d}x^\nu \).
All thermodynamical quantities, in particular the pressure \( p \), the
enthalpy per unit volume \( w = e + p \), the internal energy density
\( e \) and the temperature \( T \) are measured in the proper frame of
the fluid.  The internal energy density \( e \) constructed in this way
contains the rest mass energy density \( \rho c^2 \) and the ``purely''
thermodynamical energy density \( \varepsilon \), that is

\begin{equation}
  e = \rho c^2 + \varepsilon,
\label{eq.3}
\end{equation}

\noindent where \( \rho \) represents the proper mass density
related to the proper particle number density \( n \) and average mass 
\( m \) per particle  by the relation

\begin{equation}
  \rho = n m.
\label{eq.3.1}
\end{equation}

 In the absence of sources and sinks of particles, the particle number
density  \( n \) satisfies the continuity equation

\begin{equation}
  \frac{ \partial n^\alpha }{ \partial x^\alpha } = 0,
\label{eq.4}
\end{equation}

\noindent where \( n^\alpha \) is the particle number 
density four--vector defined by \( n^\alpha := n u^\alpha \). 

  For one dimensional flow, equation \eqref{eq.1} can be rewritten as two
equations, the equation of conservation of energy (\( \partial T^{0\alpha}
/ \partial x^\alpha = 0 \)) and the equation of conservation of momentum
(\( \partial T^{1\alpha} / \partial x^\alpha = 0 \)) respectively as

\begin{align}
  \frac{1}{c} \frac{ \partial }{ \partial t } \left( \frac{ e +\beta^{2} p }{
    1 - \beta^{2} } \right) + \frac{ 1 }{ r^k } \frac{ \partial }{ \partial r }
    \left[ r^{k} \beta \frac{ p +  e }{ 1 - \beta^{2} } \right ] = 0,
						\label{eq.7}\\
  \frac{1}{c} \frac{ \partial }{ \partial t }\left( \beta \frac{p + e}{1 - \beta^2}\right)
    + \frac{1}{r^k} \frac{ \partial }{ \partial r } \left[ r^{k} \beta^2 
    \frac{ p +  e}{1 - \beta^2 }\right ] + \frac{ \partial p }{ \partial r } 
    = 0. 
						\label{eq.8}
\end{align}

\noindent Planar, cylindrical and spherical symmetric flows
are described by these equations when \( k = 0,\ 1 \text{
and } 2 \) respectively. 

  The relativistic theory of shock waves was first investigated
by \citet{taub48,taub78} and is well described by \citet{daufm}.
Traditionally the analysis is made on a system of reference in
which the shock wave is at rest.  In a more general approach,
\citet{mckee-colgate73} wrote the necessary equations that describe
a shock wave on any inertial system of reference.  For the case of an
imploding shock wave, it is convenient to choose a system of reference
in which the un--shocked gas is at rest.  In this particular system of
reference the jump conditions across a strong shock wave are simple and
take the form \citep{mckee-colgate73,blandford76}

\begin{equation}
  \frac{ e }{ \rho } = \frac{ \gamma }{ \rho_1 } \left[ e_1 + p_1
    \frac{ \beta }{ \beta_\text{s} }\right],
\label{eq.10}
\end{equation}

\begin{equation}
  \frac{ \rho }{ \rho_1 } = \gamma \frac{ \kappa }{ \kappa - 1 } + 
    \frac{ 1 }{ \kappa - 1 },
\label{eq.11}
\end{equation}

\begin{equation}
 \Gamma = \frac{ \gamma }{ \sqrt{ 1- 2 \left( \kappa - 1 \right) /
   \kappa } }.
\label{eq.12}
\end{equation}

\noindent In the previous equations and in what follows, quantities
without subindexes and those with subindex \( 1 \) label post--shocked and
pre--shocked hydrodynamical quantities respectively. The Lorentz factor
of the post--shocked flow and that of the shock wave are represented by \(
\gamma \) and \( \Gamma \) respectively.  The corresponding values of the
velocity measured in units of the speed of light are \( \beta \) and \(
\beta_\text{s} \).  On both sides of the discontinuity, 
a Bondi--Wheeler equation of state with constant index \( \kappa \) given by

\begin{equation}
  p = \left( \kappa - 1 \right) e,
\label{eq.12.1}
\end{equation}

\noindent has been assumed.

\section{Self--similar equations and boundary conditions}
\label{self-similar-equations}

  Let us consider a strong spherical shock wave converging to a centre.
The pre-shocked material is assumed to have a constant density \(
\rho_1 \).  We analyse the problem in a state such that the radius \(
R \) of the shock wave is much smaller than the initial radius \( R_0 \)
it had when it was produced.  Under these circumstances, as it happens for
the non--relativistic case \citep{daufm,stanyukovich,zeldovichraizer},
the flow is largely independent of the specific initial conditions.
One can thus imagine a spherical piston that pushes the flow to the
centre from very large distances in such a way that it produces a strong
shock wave.  That is,

\begin{equation}
  \frac{ p }{ n } \gg \frac{ p_1 }{ n_1 }.
\label{eq.20}
\end{equation}

\noindent This expression combined with equations~\eqref{eq.10} and
\eqref{eq.11} for a gas with the equation of state~\eqref{eq.12.1} 
implies that

\begin{equation}
  \gamma - 1 \gg \frac{ 1 }{ \kappa - 1 } \, \frac{ p_1 }{ c^2 \rho_1 }
    \left(1 + \frac{ \kappa_1 }{ \kappa_1 - 1 } \, \frac{ p_1 }{ c^2
    \rho_1 } \right)^{-1},
\label{eq.21}
\end{equation}

\noindent where \( \kappa_1 \) and \( \kappa \) are the constant indexes
for the pre--shocked and post--shocked gas respectively.  The Lorentz
factor \( \gamma \) for the shocked gas that satisfies the previous
inequality is so large that the post--shocked flow velocity is comparable
to that of light.

  In order to simplify the problem, let us assume that the particles
that constitute the shocked--gas are themselves relativistic, so that
they obey an ultrarelativistic equation of state with \( \kappa = 4/3 \)
in equation~\eqref{eq.12.1}.  That is,

\begin{equation}
   p = \frac{1}{3} e.
\label{eq.22}
\end{equation}

\noindent  The shocked gas with this equation of state satisfies the
following jump velocity condition

\begin{equation}
   \gamma^2 = \frac{1}{2}\Gamma^2, 
\label{eq.23}
\end{equation}

\noindent according to equation~\eqref{eq.12}.  From this, it is clear that
inequality~\eqref{eq.21} imposes the condition \( \Gamma^2 \gg 1 \)
for a strong shock wave. This justifies the restriction of our further
analysis to zeroth order quantities in  \( \Gamma^2 \) and \( \gamma^2 \). 

  The density jump condition given by equation~\eqref{eq.11} takes
the form

\begin{equation*}
  \tilde{n} = 4 \gamma^2 n_1,
\end{equation*}

\noindent in which \( \tilde{n} \) represents the particle number density of
the shocked gas as measured with respect to the pre--shocked gas, i.e. \(
\tilde{n} = \gamma n \).  In terms of the Lorentz factor \( \Gamma \) 
of the shock wave, this last equation becomes

\begin{equation}
  \tilde{n} = 2 \Gamma^2 n_1.
\label{eq.24}
\end{equation}

  Finally, the energy jump condition \eqref{eq.10} is given by

\begin{displaymath}
  e = \frac{n}{n_1} \gamma \omega_1 = 2\Gamma^2 \omega_1.
\end{displaymath}

\noindent In terms of the pressure \( p \), the previous equation is

\begin{equation}
  p = \frac{2}{3}\Gamma^2 \omega_1,
\label{eq.25}
\end{equation}     

\noindent where \( \omega_1 = \rho_1 c^2 \) for a cold pre-shocked
gas and \( \omega_1 = 4 p_1 \) for an ultrarelativistic one.  The
three conditions given by equations~\eqref{eq.23}, \eqref{eq.24} and
\eqref{eq.25} represent post--shocked boundary conditions of the problem.
These conditions are exactly the same ones obtained by \citet{blandford76}
for the case of a relativistic explosive shock wave.

  Let us now write the equations of motion for the relativistic flow
behind the imploding spherical shock wave.  The symmetry of the
problem demands that the energy conservation equation \eqref{eq.7}
and the momentum equation \eqref{eq.8} take the form

\begin{align}
  \frac{ \mathrm{d} }{ \mathrm{d} t } \left(p \gamma^4 \right) = &
    \gamma \frac{ \partial p }{ \partial t },
    						\label{eq.26}\\
  \frac{ 1 }{ c } \frac{ \mathrm{d} }{ \mathrm{d} t } 
    \ln{\left( p^3 \gamma^4 \right)} = &
    - \frac{4}{r^2} \frac{ \partial }{ \partial r } \left( r^2 \beta \right).
						\label{eq.27}
\end{align}    

\noindent As a third equation we can either use the continuity
equation, or  the entropy conservation equation which is obtained from
equations~\eqref{eq.26}, \eqref{eq.27} and \eqref{eq.4}.  We select here
the last one, which for an ultrarelativistic gas is simply given by
\citep{daufm}

\begin{equation}
  \frac{ \mathrm{d} }{ \mathrm{d} t } \left( \frac{ p }{ n^{4/3} } \right) = 0. 
\label{eq.28}
\end{equation}

  Equations \eqref{eq.26}, \eqref{eq.27} and \eqref{eq.28} together with
the boundary conditions  \eqref{eq.23}, \eqref{eq.24} and \eqref{eq.25}
completely determine the mathematical problem.

\section{Self--similar solution}
\label{self-similar-solution}

  To find a self--similar variable for the problem, we recall that for
the case of a relativistic strong explosion \citep{blandford76}, if \( R
\) represents the radius of the shock measured from the origin, then the
ratio \( R / \Gamma^2  \) is a characteristic parameter of the problem.
With this ratio, \citet{blandford76} built a similarity variable for
the problem.  In the case of an imploding shock wave, none of the arguments
in favour of the characteristic ratio \( R / \Gamma^2 \) are valid and
instead we must construct the similarity variable by other means.
The Appendix shows an alternative way in which the similarity variable 
obtained by \citeauthor{blandford76} can be calculated.  The advantage of 
this approach is that we can generalise the method for the
case we are dealing with in this article.

  The similarity variable for the case of a strong relativistic implosion
is found as follows.  The radius of the shock wave \( R \) converges to
the origin as time increases, so that its velocity \( \mathrm{d} R /
\mathrm{d} t \) must be negative.  Thus,

\begin{equation}
  \frac{ 1 }{ c } \frac{ \mathrm{d} R}{ \mathrm{d} t} = - \left[ 1 -
   \frac{ 1 }{ \Gamma^2 }\right]^{ 1 / 2 } \approx - \left[ 1 -  \frac{
   1 }{ 2 \Gamma^2 } \right],
\label{eq.29}
\end{equation}

\noindent In order to integrate this equation, we must know the time
dependence of the Lorentz factor \( \Gamma \). For the case of a strong
relativistic explosion, this is given by equation \eqref{eq.A3}.  We can
use a similar time power law dependence by taking into consideration
the fact that the shock wave accelerates as it converges to a point.
In other words, the Lorentz factor \( \Gamma \) must increase as time
advances, that is

\begin{equation}
   \Gamma^{2} = A \left( - t \right)^{ -m },
\label{eq.30}
\end{equation}

\noindent for times \( t \leq 0 \).  The value of the constant factor
\( A \) is determined by the specific initial conditions of the
problem and \( m \) is the similarity index.  The time interval \(
t \leq 0 \) has been chosen in complete analogy to the Newtonian case
\citep[cf.][]{daufm,zeldovichraizer}.  Under these circumstances, the
time \( t = 0 \) corresponds to a time in which the shock wave collapses
to the origin, i.e.  \( R = 0 \).

  Substitution of equation \eqref{eq.30} on relation \eqref{eq.29}
yields the integration

\begin{equation}
  R = c \left| t \right| \left[ 1 - \frac{ 1 }{ 2 } \frac{ 1 }{ (m + 1)
    \Gamma^2 } \right].
\label{eq.31}
\end{equation}

\noindent In exactly the same form in which we built the similarity
variable for the case of a strong relativistic explosion (cf. Appendix),
let us choose here the similarity variable \( \eta \) as

\begin{align*}
  \eta =& 1 + 2 ( m + 1 ) \Gamma^2  \left( 1 - r / R \right),
  							\\
       =& 1 + 2 ( m + 1 ) \Gamma^2 \left[ 1 + \frac{ r }{ c t }
        \frac{ 1 }{ \left( 1-  1 / 2 ( m + 1 ) \Gamma^2 \right)}
	\right].
\end{align*}

\noindent  Thus, to order \( \textrm{O}( \Gamma^{-2} ) \), this similarity
variable takes the form

\begin{equation}
  \eta = \left( 1 + \frac{ r }{ c t } \right) \left[ 1 + 2 ( m + 1) 
    \Gamma^2 \right].
\label{eq.32}
\end{equation}

\noindent  The problem we are concerned is such that the relevant region
\( r\geq R \) corresponds to the values \( 1 \geq \eta \geq -\infty \)
of the similarity variable \( \eta \).

  For this kind of  self-similar description, Buckingham's theorem
of dimensional analysis \citep{sedov} motivated by the boundary conditions
given by equations \eqref{eq.23}, \eqref{eq.24} and \eqref{eq.25},
demands the existence of three dimensionless hydrodynamical functions \(
f(\eta) \), \( g( \eta ) \)  and \( h( \eta ) \) given by

\begin{align}
   p  =& \frac{ 2 }{ 3 } \omega_{ 1 } \Gamma^2 f( \eta ), 
   						\label{eq.33}\\ 
   \tilde{n} =& 2 n_1 \Gamma^2  h( \eta ), 
   						\label{eq.34}\\
   \gamma^2 =& \frac{ 1 }{ 2 } \Gamma^2 g( \eta ). 
   						\label{eq.35}
\end{align}

\noindent These dimensionless functions should be analytical and 
satisfy the following boundary condition at the shock wave

\begin{equation}
  f( \eta = 1 ) = g( \eta = 1 ) =  h( \eta = 1 )  =  1.
\label{eq.36}
\end{equation}

\noindent The introduction of the similarity variable \( \eta \) means
that the hydrodynamical functions change their independent variables
to \( \Gamma^2 \) and \( \eta \), instead of \( r \) and \( t \).
The derivatives of \( r \) and \( t \) in terms of these new variables
are thus given by

\begin{align}
   \frac{ \partial }{ \partial \ln t } =& - m \frac{ \partial }{ \partial \ln
     \Gamma^2 }  +  \left[ 1 + ( m + 1 )( 2 \Gamma^2 - \eta ) \right] \frac{
     \partial }{ \partial \eta},
							\label{eq.37} \\
   c t \frac{ \partial }{ \partial r} =& \left[ 1 + 2 ( m + 1 ) 
     \Gamma^2 \right] \frac{ \partial }{ \partial \eta}. 
     							\label{eq.38}
    \end{align}

\noindent Using these two equations, the total time derivative \(
\mathrm{d} / \mathrm{d} t = \partial / \partial t + v \partial / \partial
r \) takes the form

\begin{equation}
  \frac{ \mathrm{d} }{ \mathrm{d} \ln t} = - m \frac{ \partial }{ \partial 
    \ln \Gamma^2 } + ( m + 1 ) \left( \frac{ 2 }{ g } - \eta \right) 
    \frac{ \partial }{ \partial \eta }.
\label{eq.39}
\end{equation}

  \noindent With the aid of equations \eqref{eq.33}, \eqref{eq.34} and
\eqref{eq.35} we can rewrite  equations \eqref{eq.26} and \eqref{eq.27}
in terms of \( \Gamma^2 \) and \( \eta \).  In fact,  to  \( \textrm{O}(
\Gamma^{-2} ) \) we obtain

\begin{gather}
  - (m + 1) ( 2 + g \eta ) \frac{ f' }{ f g } + 4 ( m + 1 )( 2 - g \eta ) 
    \frac{ g' }{ g^2 } = 3 m,
  						\label{eq.40}\\
  \frac{ f' }{ f g } \left[ ( m + 1 ) \left( 8 - 2 g \eta \right) \right] 
    - 8 (  m + 1 ) \frac{ g' }{ g^2 } = 2 m - 8,
    						\label{eq.41}
\end{gather}

\noindent where the prime denotes derivative with respect to \( \eta \).
These two equations can be rewritten in matrix form as

\begin{equation}
  \begin{bmatrix}
    -( 2+ g \eta ) 	& 	2 ( 2 - g \eta )	&  \\
    8- 2g\eta 		& 	- 8 			&
  \end{bmatrix}
  \begin{bmatrix}
    f'/ f g 		&   \\
    g'/ g^2		&
  \end{bmatrix} = 
  \begin{bmatrix}
    3m  / ( m + 1 ) 		&   \\
    ( 2 m - 8 ) / ( m + 1 ) 	&
  \end{bmatrix}, 
\label{eq.42}
\end{equation}
\noindent which implies, according to Cramer's rule, that 

\begin{align}
  \frac{ 1 }{ f g } \frac{ \mathrm{d} f}{ \mathrm{d} \eta} = 
    \frac{ 8 ( m - 1 ) + g \eta ( 4 - m ) }{ ( m + 1 ) \left[ - 4 + 8 g 
    \eta - g^2 \eta^2 \right]}, 
    						\label{eq.43}\\
  \frac{ 1 }{ g^2 } \frac{ \mathrm{d} g }{ \mathrm{d} \eta } = 
    \frac{ 4 - 7 m + g \eta ( 2 + m ) }{ ( m + 1 ) \left[ - 4 + 8 g 
    \eta - g^2 \eta^2 \right] }. 
    						\label{eq.44}
\end{align}

\noindent Equations \eqref{eq.43} and \eqref{eq.44} are accompanied by
the boundary conditions \eqref{eq.36} for \( f( \eta ) \)  and \( g(
\eta ) \).

  The unique value for the similarity index \( m \) and the
uniqueness of the solution are determined by the following physical
consideration, analogous to the one presented in the Newtonian case
\citep[cf.][]{stanyukovich}.

  For a given time, the values of the energy density and velocity of a
particular fluid element must decrease as the radius \( r \) increases.
For the case of an ultrarelativistic gas, these quantities can be written
in terms of the pressure \( p \) and the Lorentz factor \( \gamma \)
of the flow respectively.  Thus, a necessary condition for the flow that
guarantees the decrease of these quantities is given by

\begin{displaymath}
  \frac{ \partial p }{ \partial r } < 0, \qquad \quad 
    \frac{ \partial \gamma^2
    }{ \partial r } <  0, \qquad \quad \forall \; r \; \geq \; R.
\end{displaymath}

\noindent In terms of dimensionless quantities, the previous conditions
are

\begin{equation}
  \frac{ \mathrm{d} f( \eta ) }{ \mathrm{d} \eta } > 0,
    \qquad \quad \frac{ \mathrm{d} g( \eta ) }{ \mathrm{d} \eta } > 0,
    \qquad \quad \forall \; \eta \; \leq \; 1.
\label{eq.45}
\end{equation}

  The integral functions \( f( \eta ) \) and \( g( \eta ) \) depend
on the choice of the parameter \( m \) and so, a family of integral
curves can be found for different self--similar flows.  However, with
the restrictions imposed by inequalities \eqref{eq.45}, it is possible
to integrate the differential equations \eqref{eq.43} and \eqref{eq.44}
in a unique form.  This happens because the conditions \eqref{eq.45}
impose a single value for the similarity index \( m \).

  The derivative \( \mathrm{d} g( \eta ) / \mathrm{d} \eta \) in  equation
\eqref{eq.44}  has a positive value at the surface of the shock wave, i.e.
when \( \eta = g = 1 \).  In order to guarantee a positive value for all
\( \eta < 1 \), \( g( \eta ) \) must avoid extrema.  In the same form,
the continuity of the function \( g( \eta ) \) means that the denominator
on the right hand side of equation \eqref{eq.44} should not vanish.
To overcome these problems, we restrict the integral curve to pass through
the singular point (\(\eta*,\, g(\eta*) \)) for which the numerator and
denominator on equation \eqref{eq.44} vanish simultaneously.

  The denominator on equation \eqref{eq.44} vanishes when the following
condition is satisfied

\begin{equation}
  g \eta = 4\pm 2 \sqrt{ 3 },
\label{eq.46}
\end{equation}

\noindent which represents a hyperbola on the \( g \eta \) plane.
The numerator of the same equation has a null value on the hyperbola
 
\begin{equation}
  4 - 7m + g \eta (2 + m) = 0,
\label{eq.47}
\end{equation}

\noindent The point (\(\eta*, g( \eta* ) \))  where both hyperbolas intersect is
calculated by direct substitution of the value \( g \eta \) on equation
\eqref{eq.47}, choosing the negative sign so that the number \( m \)
remains positive. This fixes a unique value for the similarity index \(
m \) given by

\begin{equation}
  m = 12 \sqrt{3} - 20 = 0.78460969. 
\label{eq.48}
\end{equation}

  If we now follow the same procedure for the derivative \( \mathrm{d}
f( \eta ) / \mathrm{d} \eta   \) in equation \eqref{eq.43}, we obtain
the same value for the similarity index \( m \) as the one given by the
previous equation.

\begin{figure}
  \begin{center}
    \includegraphics[scale=0.9]{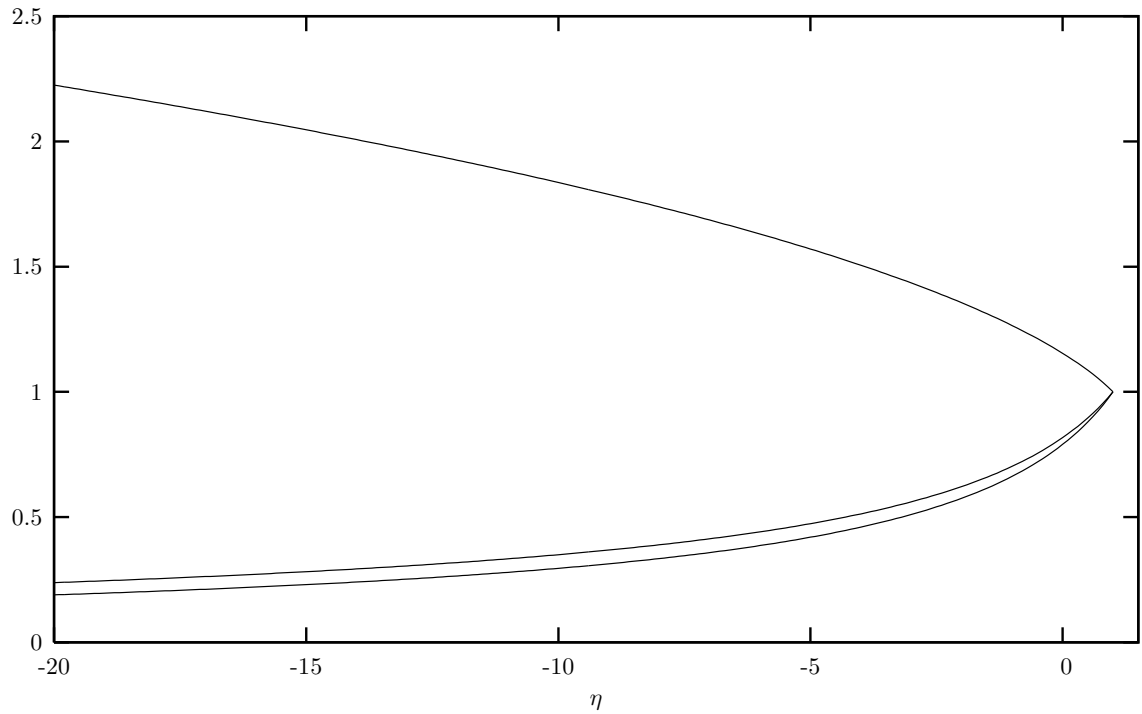}
  \end{center}
  \caption[Integrals of dimensionless functions]{ The figure shows,
           from bottom to top, the integral curves of the dimensionless
	   pressure \( f(\eta) \), squared Lorentz factor \( g(\eta) \) and
	   particle number density \( h(\eta) \) as functions of the
	   similarity variable \( \eta \).}
\label{fig.0}
\end{figure}

  The value of the similarity index given by equation \eqref{eq.48}
guarantees that  pressure and velocity gradients have negative values for
the shocked gas.  Using a 4th rank Runge--Kutta method we have calculated
the integral curves for \( g( \eta ) \) and for \(f(\eta)\).  These results
are plotted in Fig.~\ref{fig.0}.

  Let us now obtain a differential equation for the function \( h( \eta )
\).  This is obtained by direct substitution of the differential equations
\eqref{eq.43} and \eqref{eq.44} in equation \eqref{eq.28}, that is

\begin{equation}
  \frac{ 1 }{ g h } \frac{ \mathrm{d} h }{ \mathrm{d} \eta } =
    \frac{ 2 \left( 8 - 9m \right) + 2 g \eta \left(
    5m - 6 \right) - g^2 \eta^2 \left( m -  2 \right) }{
    ( 2 - g \eta ) ( m + 1 ) \left[ -4 + 8 g \eta - g^2 \eta^2 \right]}.
\label{eq.49}
\end{equation}

\noindent The integral curve \( h( \eta ) \) of this equation shown in
Figure~\ref{fig.0} was calculated with the already integrated values
of \( g( \eta ) \) and the unique value of \( m \) given by equation
\eqref{eq.48}.  The number \( m \) is such that when the numerator on the
right hand side of the previous equation vanishes, the numerator also
does so, avoiding singularities and extrema in the range \( \eta \leq
1 \).  Furthermore, the derivative \( \mathrm{d} h / \mathrm{d} \eta \)
evaluated on \( \eta = 1 \) has a negative value.  Thus, the particle
number density \( \tilde{n} \) as presented by relation \eqref{eq.34}
grows monotonically as \( r \) increases for fixed values of time \(
t \), i.e.

\begin{equation}
  \frac{ \mathrm{d} h(\eta) }{ \mathrm{d} \eta } < 0.
\label{eq.49b}
\end{equation}

\noindent This peculiar situation also occurs for the non--relativistic
imploding shock wave and will be discussed in more detail in the next
section.

  With the integral functions \( f( \eta ) \), \( g( \eta ) \) and \(
h( \eta ) \) calculated in this section it is then possible to calculate
the dimensional hydrodynamical quantities using equations \eqref{eq.33},
\eqref{eq.34} and \eqref{eq.35}.

\section{Discussion}
\label{discussion}

  The self--similar solution found in this article shows that some of
the values of the hydrodynamical quantities decrease in a given fluid
element as it moves behind the shock wave.  For example, the variation
of the pressure in time as a fluid element moves behind the shock wave
satisfies the following inequality

\begin{align}
  \frac{ \mathrm{d} p }{ \mathrm{d} t } =& - \frac{ 1 }{ \left| t 
    \right| } \frac{ \mathrm{d} p }{ \mathrm{d}
    \ln{t} },
							\notag \\
    =& \frac{ 1 }{ \left| t \right| } \left\{ - m \frac{ \partial
    }{ \partial \ln \Gamma^2 } + \left( m + 1 \right) \left( \frac{ 2 }{
    g } - \eta \right) \frac{ \partial }{ \partial \eta } \right\} p,
  							\notag \\
  =& - \frac{ 1 }{ \left| t \right| } \left\{ - m p  + p \left( m + 1
    \right) \left(  2 - g \eta \right) \left[ \frac{ 1 }{ g f } \frac{
    \partial f }{ \partial \eta } \right] \right\},
  							\notag \\
  <& - \frac{ 1 }{ \left| t \right| } \left\{ - ( m + 1 ) p + 
    p \left( m + 1 \right) \left( 2 - g \eta \right)
    \left[ \frac{ 1 }{ g f } \frac{ \partial f }{ \partial \eta } 
    \right] \right\},
  							\notag \\
  <& - \frac{ 1 }{ \left| t \right| } \left\{ p 
    \left( m + 1 \right) \left( 1 - g \eta \right)
    \left[ \frac{ 1 }{ g f } \frac{ \partial f }{ \partial \eta } 
    \right] \right\}.
    							\label{eq.50}
\end{align}

\noindent The solution obtained in the previous section
satisfies the condition \( g \eta \leq 1 \), for \( \eta \leq
1 \) (cf.  Figure~\ref{fig.0}). From this and the condition in
equation~\eqref{eq.45}, it follows that the total time derivative \(
\mathrm{d} p / \mathrm{d} t \) has a negative value for any fluid element
behind the shock wave.  Using similar arguments it is easy to prove
that, as the shocked gas particles move away from the shock wave, the
Lorentz factor \( \gamma \) decreases and the particle number density \(
\tilde{n} \) increases.  These results combined with the fact that \(
\tilde{n} = \gamma n \) imply that the proper particle number density \(
n \) increases as the fluid particles move away from the shock wave,
i.e. \( \mathrm{d} n / \mathrm{d} t > 0 \).  This result is exactly
what is obtained in the non--relativistic case \citep{zeldovichraizer}.
The particle number density increases behind the shock and at sufficiently
large distances behind the shock it must converge to a finite value.

  Self--similar solutions of the second type (see e.g.
\citet{zeldovichraizer}) such as the one found in this article satisfy
two basic properties.  The first one is the fact that the integral curve
passes through a singular point (for a specific value of the similarity
index).  The second one is the existence of a curve \( r(t) \) on the
\(r\)--\(t\) plane which corresponds to the singular point \( \eta_* \)
and is itself a C\(_-\) characteristic bounding the region of influence.
In order to verify this last statement for the self--similar solution
found above, we proceed as follows.  Let \( \eta_* = \text{const.}
\) correspond to a value of the similarity variable evaluated at the
singular point.  Using equations~\eqref{eq.30} and \eqref{eq.32} it is
then straightforward to calculate the differential equation satisfied
by the curve \( r_*(t) \) evaluated at the singular point:

\begin{equation}
  \frac{ \mathrm{d} r_* }{ \mathrm{d} t } = -c + \frac{ \eta_* c }{ 2
    \Gamma^2 },
\label{eq.53}
\end{equation}

\noindent at \( \textrm{O}( \Gamma^{-2} ) \).  On the other hand,
at the singular point, equation~\eqref{eq.44} implies that the product
\( \eta_* g_* = 2 / \left( 2 + \sqrt{3} \right) \) for the case \( 0
\leq \eta \leq 1 \) which corresponds to the light causal region \( R \leq r
\leq c |t| \), according to equations~\eqref{eq.31} and \eqref{eq.32}.
With this, equation~\eqref{eq.53} can be rewritten as 

\begin{equation}
  \frac{ \mathrm{d} r_* }{ \mathrm{d} t } = -c + \frac{ c }{ 2 \left( 2 +
  \sqrt{3} \right) \gamma_*^2 }.
\label{eq.54}
\end{equation}

  The curve \( r(t) \) that describes the trajectory of any C\(_- \)
characteristic on the \(r\)--\(t\) plane is described by the following
differential equation \citep{taub48,taub78}

\begin{equation}
  \frac{ \mathrm{d} r }{ \mathrm{d} t } = -c \, \frac{ \alpha - \beta }{ 1
    - \alpha \, \beta },
\label{eq.55}
\end{equation}

\noindent where \( \alpha \) and \( \beta \) are the sound speed and
the velocity of the flow measured in units of the speed of light \(
c \) respectively.  For the self--similar solution we are dealing with
in this article, \( \beta \approx -1 + 1 / 2\gamma^2 \).  This implies
that at \( \textrm{O}( \gamma^{-2} ) \), equation~\eqref{eq.55} takes
the following form

\begin{equation}
  \frac{ \mathrm{d} r }{ \mathrm{d} t } = -c + \frac{ c \left( 1 - \alpha 
    \right) }{ 2 \gamma^2 \left( 1 + \alpha \right) }.
\label{eq.56}
\end{equation}

\noindent Now, using the equation of state~\eqref{eq.22}, it follows
that the value of the velocity of sound in units of the speed of light
is given by \( \alpha = 1 / \sqrt{3} \).  So, equation~\eqref{eq.56}
has exactly the same form as equation~\eqref{eq.54}.  This means that
the curve \( r_*(t) \) that corresponds to the singular value \( \eta_*
\) is itself a C\(_-\) characteristic.  In what follows we label this
particular characteristic as \( C_-^* \). From the above results,
an important condition about the causality of the flow behind the
imploding shock wave is obtained.  Since the curve \( r_*(t) \) that
corresponds to the singular point \( \eta_* \) does not intersect the
imploding shock wave (except at the origin of the \( r \)--\( t \)
plane) and because the C\(_-\) characteristics do not intersect one
another, there is a region near the shock wave for which all C\(_-\)
characteristics intersect the shock wave.  Beyond the curve \( r_*(t) \),
the C\(_-\) characteristics do not intersect the imploding shock wave.
This means that the flow farther away from the singular C\(_-^*\)
characteristic does not have any causal influence on the shock wave.
These physical consequences on the causality are exactly the same as
the ones that occur for the non--relativistic imploding shock wave.

  The energy contained inside a gas shell that moves in a self--similar
manner can be calculated as follows.  First, let us write down the value
for the energy \( E \) contained on a shell of shocked gas.  For an
ultrarelativistic gas, the energy density \( T^{00} \) measured on the
system of reference in which the pre--shocked gas is at rest is given
by \( p ( 4 \gamma^2 - 1 ) \).  Thus, to \( \textrm{O}( \gamma^{-2} )
\) inside the shell limited by the radius \( R_0(t) \) and \( R_1(t) \)
the energy is given by

\begin{equation}
  E( R_0( t ), \, R_1( t ),\, t ) = 16 \pi \int_{ R_0 }^{ R_1 }{ p \gamma^2
    r^2\, \mathrm{d} r }.
\label{eq.50.1}
\end{equation}

  When the limits of integration are taken from the radii \( r = R \) to
\( r_* = c |t| \left( 1 - \eta_* / [ 1 + 2 ( m + 1 ) \Gamma^2 ] \right) \),
i.e. the self--similar region of the flow, we obtain

\begin{align}
  E =& 8 \pi \omega_1 \Gamma^4  \left| t \right|^3 \left( 1 + 2 ( m +
    1 ) \Gamma^2 \right)^{-1} \int_{ 1 }^{ \eta_* } f g \left( 1 - \frac{
    \eta }{ 1 + 2 ( m + 1 ) \Gamma^2 } \right)^2 \, \mathrm{d} \eta,
  							\notag \\
  \approx & 8 \pi \omega_1 \Gamma^2 \left| t \right|^3 \Xi( \eta_* ).
\label{eq.51}
\end{align}

\noindent In these relations, the quantity \( \Xi( \eta_* ) \) is a constant
and all quantities were approximated to \( \textrm{O}( \Gamma^{-2} ) \).
Substitution of equation \eqref{eq.30} in equation \eqref{eq.51} yields

\begin{equation}
  E= 8 \pi \omega_1 A \left|t \right|^{3 - m} \Xi(\eta_*).
\label{eq.52}
\end{equation}

\noindent  This equation, together with the value of the similarity index
given by equation~\eqref{eq.48}, means that the energy contained inside
the shell is proportional to a positive power of time.  In other words,
the energy inside this self--similar shell tends to zero as the imploding
shock wave collapses to the origin, which occurs when \( t \rightarrow
0 \).  As the shock wave converges to a point, energy becomes
concentrated right behind the shock front. However the dimensions of the
self--similar region decrease and the previous analysis shows that
the energy concentrated within this region also decreases.

\begin{figure}
  \begin{center}
    \includegraphics[scale=0.9]{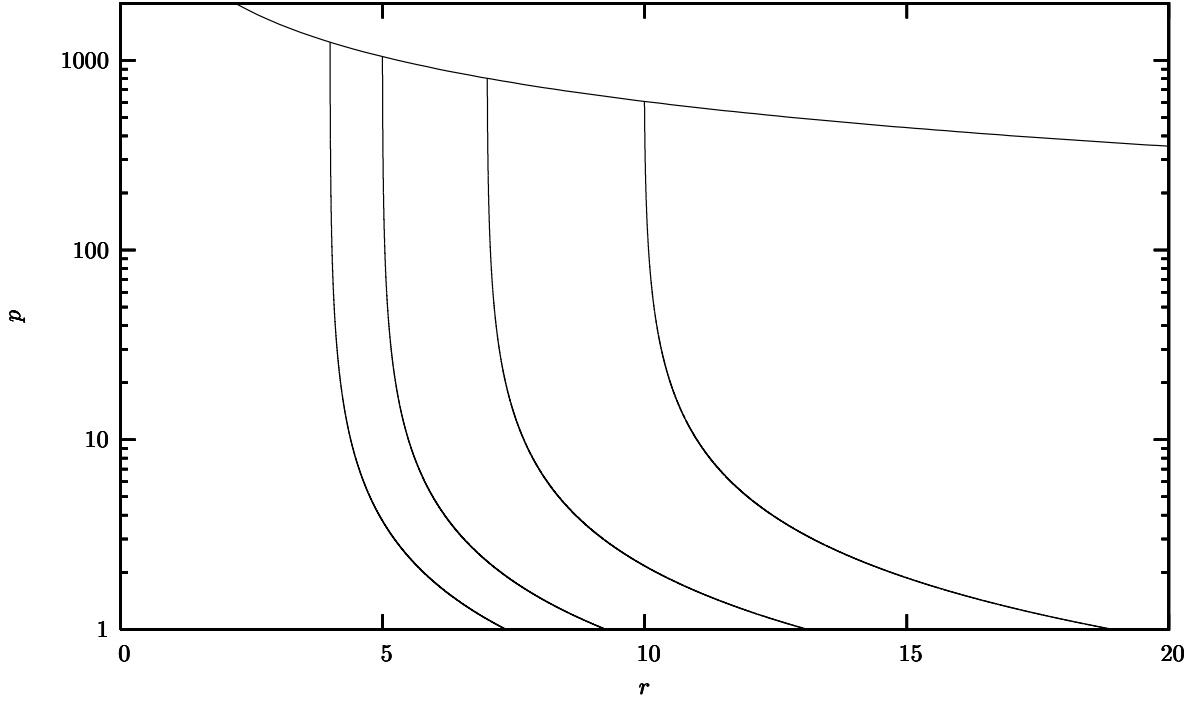}
  \end{center}
  \caption[Gas pressure profiles behind a relativistic imploding shock
	   wave.]{ The figure shows pressure \( p \) profiles behind
	   a strong relativistic imploding shock wave.	Natural units
	   for the plot were used.  The ambient medium was assummed
	   to be of uniform density, with a particle number density \(
	   n_1 = 1 \), a pressure \( p_1 = 1 \) and an enthalpy per unit
	   volume \( w_1 = 1 \).  The imploding shock wave was considered
	   to be formed at a radius \( r_0 = 100 \), where the initial
	   pressure discontinuity \( p_2 = 100 \).  From left to right,
	   the figure shows the pressure distribution as a function
	   of radius \( r \) for time values of \( 4 \), \( 5 \), \(
	   7 \) and \( 10 \) before the collapse of the shock wave.
	   The envelope for the different values of the pressure just
	   behind the shock wave is also shown.  The pressure grows
	   without limit as the shock wave converges to the origin.}
\label{fig.1}
\end{figure}

\begin{figure}
  \begin{center}
    \includegraphics[scale=0.9]{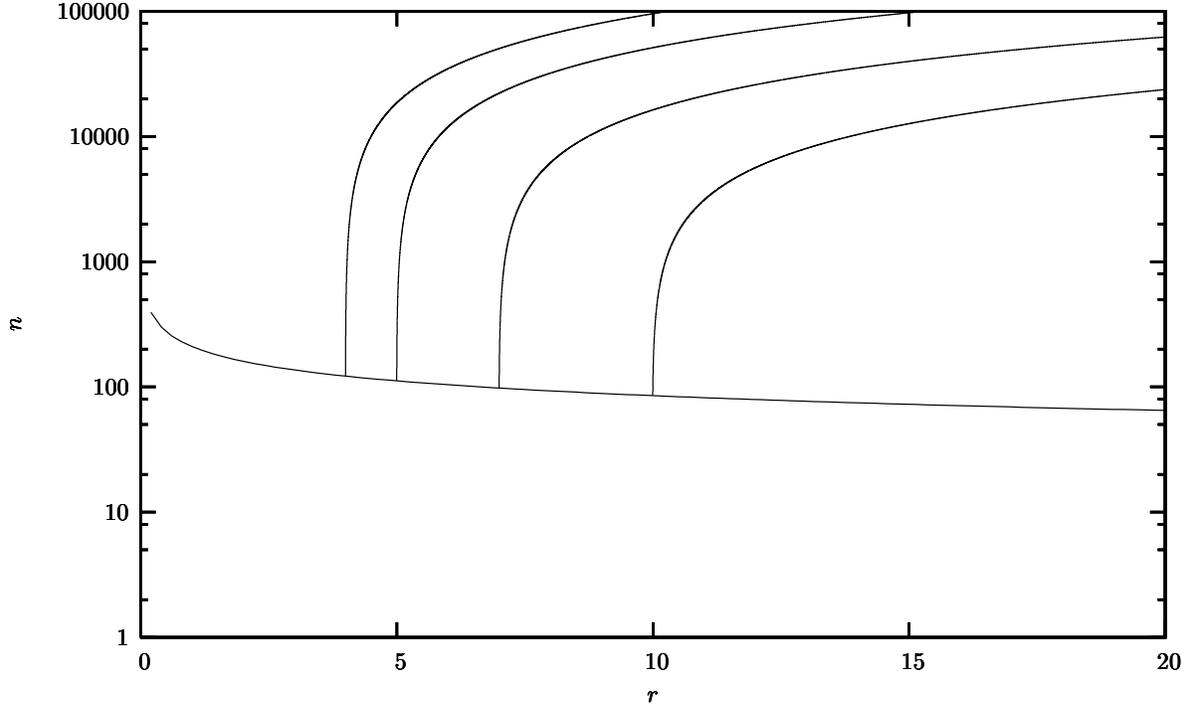}
  \end{center}
  \caption[Particle number density profiles behind a relativistic
	   imploding shock wave.]{ Particle number density \( n \)
	   profiles for the conditions mentioned in Figure~\ref{fig.1}
	   are shown in the figure, as a function of the radial distance
	   \( r \).  From left to right, the plotted curves show the
	   particle number density distribution for times of  \( 4 \),
	   \( 5 \), \( 7 \) and \( 10 \) previous to the collapse of
	   the shock wave respectively.  The envelope curve shows the
	   particle number density for the gas that flows just behind
	   the shock wave.  The particle number density \( n \) tends
	   to infinity as the shock wave reaches the origin.  Note that
	   the particle number density profiles appear to grow without
	   limit for sufficiently large radius behind the shock wave.
	   This is a false result due to the fact that the Lorentz factor
	   of the flow \( \gamma \) is such that \( \gamma^2 \lesssim
	   10 \) and the assumptions made on this article are no longer
	   valid on this regime (see text for more information on this).}
\label{fig.2}
\end{figure}

\begin{figure}
  \begin{center}
   \includegraphics[scale=0.9]{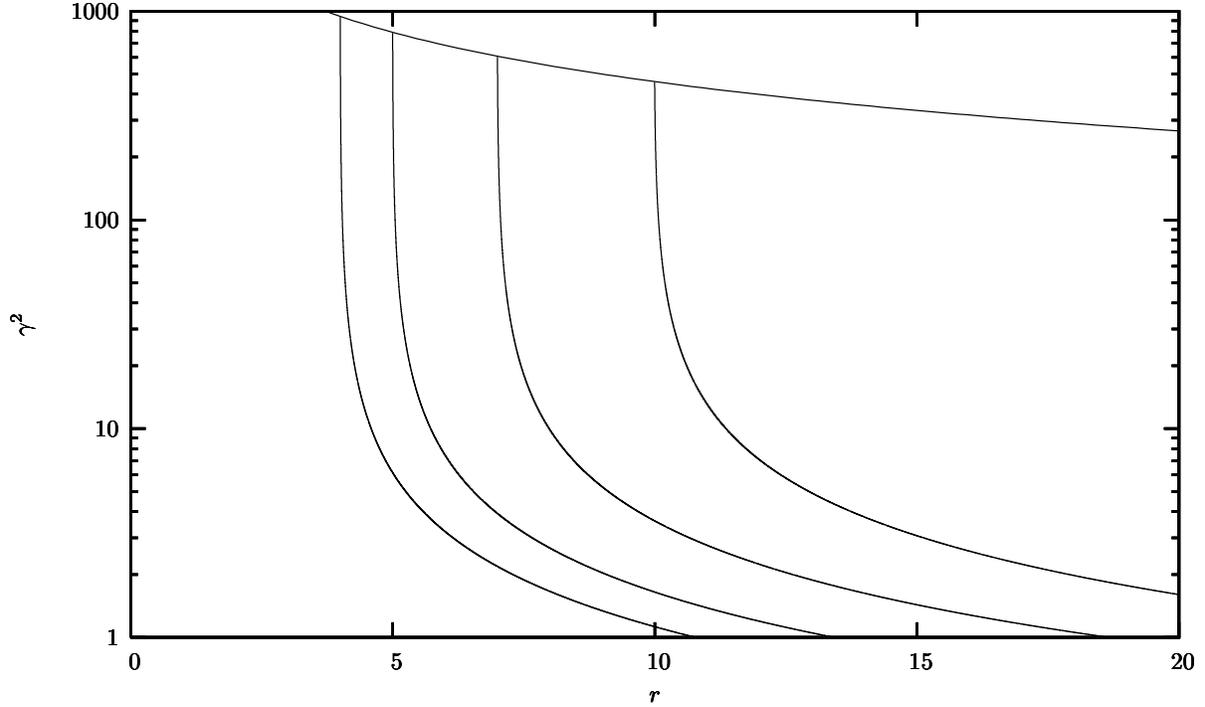}
  \end{center}
  \caption[Lorentz factor profiles for the gas behind a relativistic 
           imploding shock wave.]{ Different Lorentz factor \( \gamma \)
	   profiles as function of the radial distance \( r \) are shown 
	   in the figure.  The profiles
	   are calculated for the shocked gas behind a strong imploding
	   relativistic shock wave travelling through a constant gas medium
	   with the same conditions as the one shown in
	   Figure~\ref{fig.1}.  This plot shows, from left to right,
	   Lorentz factor values corresponding to times of \( 4 \), \( 5
	   \) \( 7 \) and  \( 10 \) before the collapse of the imploding
	   shock wave.  The envelope curve shows the values of the Lorentz
	   factor for particles that have just crossed the strong shock
	   wave.  As the plot shows, the Lorentz factor
	   diverges as the shock wave collapses to the origin.  Note that
	   the solution breaks down for sufficiently small values of \(
	   \gamma \) such that the \( \text{O}(\gamma^{-2}) \)
	   approximation is no longer valid.}
\label{fig.3}
\end{figure}

  The constant \( A \) was introduced in equation~\eqref{eq.30}.
Its value is determined by the initial conditions that generate
the shock wave, in complete analogy with the Newtonian case (cf.
\citep{zeldovichraizer}).  Thus, through a specified value of this
constant, different flows behind the imploding shock wave are generated
for different values of the initial pressure discontinuity.  Figures
\ref{fig.1} to \ref{fig.3} show examples of numerical integrations of the
equations of motion.  These solutions were calculated in natural units
such that \( c = k_\text{B} = 1 \), with \( k_\text{B} \) representing
Boltzmann's constant.  The pre--shocked uniform medium was normalised in
order to have a particle number density \( n_1 = 1 \), initial pressure \(
p_1 = 1 \) and initial enthalpy per unit volume \( w_1 = m n_1 c^2 = 1 \).
These assumptions fix uniquely the units of mass, length and time for the
specific problem.  The initial pressure discontinuity that gave rise to
the shock was assumed to have a value \( p_2 = 100 \) at a distance \(
R_0 = 100 \). The approximations made in this article break down for
values \( \gamma \lesssim 10 \).  As discussed before, the pressure
and the Lorentz factor behind the flow decrease with increasing radius.
The particle number density seems to grow without limit behind the shock
wave.  This unphysical behaviour occurs because the solutions discussed
above are valid only for \( \text{O}(\gamma^{-2}) \), and the region
where the proper particle number density converge does not satisfy this
approximation.  Only a full self--similar solution of the problem, with no
first order approximations in \( \gamma^{-2} \), will show that behaviour.

  For the particular case mentioned in this article, we have presented
the last stages before the collapse of an imploding shock wave, which
was generated by a very strong external pressure and where the complete
treatment requires a relativistic analysis.  At these final stages of the
implosion, the flow pattern is not too sensitive of the specific initial
conditions of the problem.

  Finally, we mention the regime under which the discussed solution
remains valid.  This is simply determined by the value of the
post--shocked flow velocity.  For sufficiently large distances behind
the shock wave, the velocity of the gas \( \beta \) in units of the
speed of light reduces so that its Lorentz factor \( \gamma \sim 1 \).
In this velocity regime, the approximations made on this article are no
longer valid and the relativistic self--similar solution has no meaning.

\begin{acknowledgments}
  We gratefully thank the anonymous referee who pointed out the
importance of the singular C\(^*_-\) characteristic related to the nature
of the self--similar flow of the second type discussed in this article.
J.~C. Hidalgo acknowledges financial support from Fundaci\'on UNAM and
CONACyT~(179026).  S. Mendoza gratefully acknowledges financial support
from CONACyT~(41443) and DGAPA~(IN119203-3).
\end{acknowledgments}

\appendix
\section{Relativistic similarity variable for a strong explosion}
\label{appendix}

  Let us consider a relativistic strong explosion, diverging from the
origin of coordinates.  As mentioned in the article, similarity solutions
were found by \citet{blandford76}.  Let us show in here how to obtain their
similarity variable with an alternative argument that can be extended
naturally for the case of a strong relativistic implosion.

  The rate of change of the radius \( R(t) \) of the shock wave  in units
of the speed of light is given by

\begin{equation}
  \frac{ 1 }{ c } \frac{ \mathrm{d} R }{ \mathrm{d} t } = \left( 1 -
    \frac{ 1 }{ \Gamma^2 } \right)^{ 1/ 2 },
\label{eq.A1}
\end{equation}

\noindent where \( \Gamma \) represents the Lorentz factor of the shock
wave.  A Taylor expansion of the right hand side of this equation to
\( \text{O}(\Gamma^{-2}) \) gives

\begin{equation}
 \frac{ \mathrm{d} R }{ \mathrm{d} t } = c \left( 1 - \frac{ 1 }{
   2 \Gamma^2 } \right).
\label{eq.A2}
\end{equation}

\noindent Let us consider that \( \Gamma^2 \) is a power law
function of time only, so that

\begin{equation}
  \Gamma^2 = \frac{ A }{ t^m }, 
\label{eq.A3}
\end{equation}

\noindent in which \( A \) is an unknown constant (but can be found from
the specific initial conditions of the problem) and the constant \(
m \) represents the similarity index.  Substitution of
equation \eqref{eq.A3} on \eqref{eq.A2} yields the integral

\begin{equation}
  R = c t \left( 1 - \frac{ 1 }{ m + 1 } \frac{ 1 }{ 2 \Gamma^2 }\right). 
\label{eq.A4}
\end{equation}

\noindent The product \( c t \) has dimensions of length.  Thus, we can
write the dimensionless ratio

\begin{displaymath}
  \frac{ R }{ r } := 1 - \frac{ \phi }{ 2 \Gamma^2 \left( m+ 1 \right) },
\end{displaymath}

\noindent where we have introduced a new dimensionless quantity \( \phi \).
To first order approximation, the inverse of the previous relation is
given by 

\begin{displaymath}
   \frac{ r }{ R } = 1 + \frac{ \phi }{ 2 \Gamma^2 (m + 1) },
\end{displaymath}

\noindent so that the dimensionless variable \( \phi \) is 

\begin{equation}
   \phi = \left( 1 - \frac{r}{R} \right) 2 (m + 1) \Gamma^2,
\label{eq.A5}
\end{equation}

\noindent and vanishes when $r = R$.

  Self--similar motion means that all hydrodynamical functions can be
written as the product of two dimensionless functions \( \Pi(t) \text{
and } \pi(r/R) \) \citep{sedov,zeldovichraizer}, which is satisfied by the
parameter \( \phi \) on equation~\eqref{eq.A5}.  To avoid a null value
of this variable at the shock radius, it is more convenient to take as
similarity variable the function \( \chi :=  \phi + 1 \), and not \(
\phi \), i.e.

\begin{align}
  \chi = 1 + \left( 1 - \frac{ r }{ R } \right) 2 ( m + 1 ) \Gamma^2. 
\label{eq.A6}
\end{align}

\noindent The range of this similarity variable should be taken for \(
\chi \geq 1 \).  The value \( \chi = 1 \) corresponds to the shock wave.
As described by \citet{blandford76}, the hydrodynamical equations simplify
to great extent when this variable is used.   Substitution of equations
\eqref{eq.A3} and \eqref{eq.A4} on \eqref{eq.A6} yields an expression
for \(\chi\) at \( \textrm{O}( \Gamma^{-2} ) \) given by

\begin{equation}
  \chi = \left( 1- \frac{ r }{ c \, t } \right) \left[ 1+ 2 ( m + 1 )
    \Gamma^2 \right].
\label{eq.A7}
\end{equation}


\begin{thebibliography}{14}
%
\expandafter\ifx\csname natexlab\endcsname\relax\def\natexlab#1{#1}\fi
\expandafter\ifx\csname url\endcsname\relax
  \def\url#1{{\tt #1}}\fi
\expandafter\ifx\csname urlprefix\endcsname\relax\def\urlprefix{URL }\fi
\expandafter\ifx\csname selectlanguage\endcsname\relax
  \def\selectlanguage#1{\relax}\fi

\bibitem[{{Blandford} and {McKee}(1976)}]{blandford76}
{ R.~D. {Blandford}, and  C.~F. {McKee}},
\newblock {``Fluid dynamics of relativistic blast waves,''}
\newblock {Phys. Fluids\/} {\bf 19}, 1130 (1976).

\bibitem[{{Eltgroth}(1971)}]{eltgroth71}
{ P.~G. {Eltgroth}},
\newblock {``Similarity analysis for relativistic flow in one dimension,''}
\newblock {Phys. Fluids\/} {\bf 14}, 2631 (1971).

\bibitem[{{Eltgroth}(1972)}]{eltgroth72}
{ P.~G. {Eltgroth}},
\newblock {``Nonplanar Relativistic Flow,''}
\newblock {Phys. Fluids\/}, {\bf 15}, 2140 (1972).

\bibitem[{{Guderley}(1942)}]{guderley42}
{ G. {Guderley}},
\newblock {``Starke kugelige und zylindrische Verdigtunst\"{o}sse in
  der N\"{a}he des Kugelmittelpunktes bzw. der Zilinderische,''}
\newblock {Luftfahrtforschung\/}, {\bf 19}, 302 (1942).

\bibitem[{Landau and Lifshitz(1987)}]{daufm}
{ L. Landau, and E. Lifshitz},
\newblock {\emph{Fluid Mechanics}\/} 
\newblock (Pergamon Press, London, 1987).

\bibitem[{{McKee} and {Colgate}(1973)}]{mckee-colgate73}
{ C.~R. {McKee}, and S.~A. {Colgate}},
\newblock {``Relativistic Shock Hydrodynamics,''}
\newblock {\apj\/}, {\bf 181}, 903, (1973).

\bibitem[{{Sedov}(1946)}]{sedov46}
{ L. {Sedov}}, 
\newblock {``Propagation of strong explosive waves,''}
\newblock {Prikl. Mat. Mekh.\/}, {\bf 9}, 2 (1946).

\bibitem[{{Sedov}(1993)}]{sedov}
{ L. {Sedov}},
\newblock {\emph{Similarity and Dimensional Methods in Mechanics}\/}
\newblock (CRC Press, Florida, 1993).

\bibitem[{{Stanyukovich}(1960)}]{stanyukovich}
{ K.~P. {Stanyukovich}},
\newblock {\emph{Unsteady motion of continuous media}\/}
\newblock (Pergamon Press, New York, 1960).

\bibitem[{{Taub}(1948)}]{taub48}
{ A.~H. {Taub}}, 
\newblock {``Relativistic Rankine--Hugoniot Equations,''}
\newblock {Physical Review\/}, {\bf 74}, 328 (1948).

\bibitem[{{Taub}(1978)}]{taub78}
{ A.~H. {Taub} },
\newblock {``Relativistic Fluid Mechanics,''}
\newblock {Annual Reviews of Fluid Mechanics\/}, {\bf 10}, 301 (1978).

\bibitem[{{Taylor}(1950)}]{taylor45}
{ G.~I. {Taylor}},
\newblock {``The Formation of a blast wave by a very intense explosion II. The
  atomic explosion of 1945,''}
\newblock {Proc.Roy. Soc. Ser. A\/}, {\bf 201}, 175 (1950).

\bibitem[{{Zel'dovich} and {Raizer}(2002)}]{zeldovichraizer}
{ I.~B. {Zel'dovich}, and Y.~P. {Raizer}}, 
\newblock {\emph{Physics of Shock Waves and High-Temperature Hydrodynamic
  Phenomena}\/}
\newblock (Dover Press, New York, 2002).

\end{thebibliography}
%
%

\vfill \newpage

\end{document}